\documentclass[manuscript,screen]{acmart}

\AtBeginDocument{%
  \providecommand\BibTeX{{%
    \normalfont B\kern-0.5em{\scshape i\kern-0.25em b}\kern-0.8em\TeX}}}

\setcopyright{rightsretained}
\copyrightyear{2023}
\acmYear{2023}
\acmDOI{...}

\acmConference[]{}{}
\acmBooktitle{}
\acmPrice{}
\acmISBN{...}

\begin{document}


\title{Challenges and Opportunities in Providing Small Farmers Equal Access to Wealth via Rural Credit in Brazil}


\author{Vagner Figueredo de Santana}
\email{vsantana@ibm.com}
\orcid{0003-0325-1596}
\affiliation{%
  \institution{IBM Research}
  \city{Yorktown Heights}
  \state{New York}
  \country{United States}
}

\author{Raquel Zarattini Chebabi}
\email{rchebabi@br.ibm.com}
\affiliation{%
  \institution{IBM Research}
  \city{São Paulo}
  \state{São Paulo}
  \country{Brazil}
}

\author{David Millen}
\email{dmillen@us.ibm.com}
\affiliation{%
  \institution{IBM Research}
  \city{São Paulo}
  \state{São Paulo}
  \country{Brazil}
}


\begin{abstract}
Agriculture is impacted by multiple variables such as weather, soil, crop, stocks, socioeconomic context, cultural aspects, supply and demand, just to name a few. Hence, understanding this domain and identifying challenges faced by stakeholders is hard to scale due to its highly localized nature. 
This work builds upon six months of field research and presents challenges and opportunities for stakeholders acting in the rural credit ecosystem in Brazil, highlighting how small farmers struggle to access higher values in credit.
This study combined two methods for understanding challenges and opportunities in rural credit ecosystem in Brazil: (1) a study that took place in a community of farmers in Brazil and it was based on participatory observations of their work processes and interactions of 20 informants (bank employees and farmers); (2) design thinking workshops with teams from 3 banks, counting on 15-20 participants each.
The results show that key user experience challenges are tightly connected to the heterogeneity of farmer profiles and contexts of use involving technology available, domain skills, level of education, and connectivity, among others. In addition to presenting data collected from interaction with informants and experiences resulting from active participant observation, we discuss a holistic view of how recommender systems could be used to promote better bank-farmer interactions, improve farmer experience in the whole process, and promote equitable access to loans beyond microcredit.
\end{abstract}



\begin{teaserfigure}
  \includegraphics[trim=0 400 0 1000, clip, width=\textwidth]{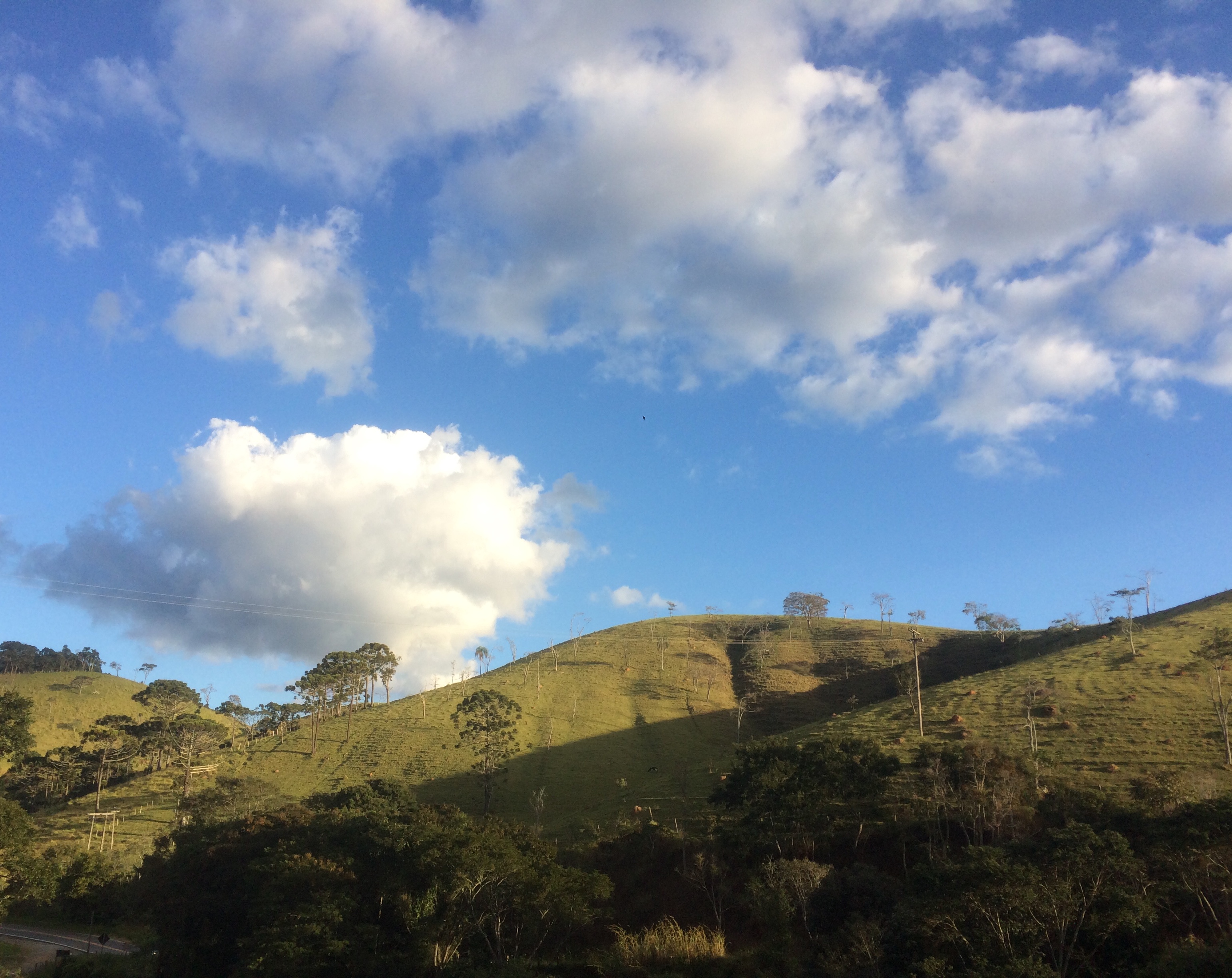}
  \caption{Photo diary picture showing the landscape of the region where the study took place, Eastern State of São Paulo, Brazil.}
  \Description{Photo diary picture showing the landscape of the region where the study took place, Eastern State of São Paulo, Brazil. It shows green mountains with trees and a blue sky in the background with a couple of small clouds.}
  \label{fig:sao_bento}
\end{teaserfigure}

\maketitle

%

\section{Introduction}

Agriculture involves business, science, or the activity of farming\footnote{https://dictionary.cambridge.org/us/dictionary/english/agriculture}. And, as it continues to play a key role in our society, there are still places where crops are managed as centuries ago, representing one of the more heterogeneous work activities still in place in the twenty-first century in terms of tools, technologies, skills, and socioeconomic contexts. This complex ecosystem\footnote{The use of the term ecosystem refers to the complex network of rural credit business processes, its multiple stakeholders, the digital systems in place, and its connection with the environment.} creates challenges and opportunities for Human-Computer Interaction (HCI) practitioners on understanding, designing, and evaluating products and services in this realm. Moreover, this intrinsically context-dependent aspect of agriculture (weather, soil, crop, stocks, socioeconomic context, supply, demand, cultural aspects, etc.) poses challenges to studies aiming at any type of generalization of research results. 
Thus, in this work, we convey this challenge by using the term \textbf{localized challenges} to refer to this context-dependent aspect, intrinsic to agriculture. Moreover, promoting agriculture is strategic for multiple economies around the globe and Brazil is not an exception. 
However, few solutions are designed having in mind adoption by small farmers, infrastructure, and cultural challenges faced in the Global South \cite{Chandra2021}.
Hence, this work discusses the role of technology and its potential to provide small farmers equal access to wealth in Brazil. 
The biggest agriculture credit initiative in Brazil is Rural Credit and it has the potential to change the lives of small farmers, but there is a gap between the credit lines offered and the credit they actually obtain; this is where this research is situated.
More details on Rural Credit in Brazil are provided next.

\subsection{Rural Credit in Brazil}

Rural Credit is the main component of Brazil's strategy to agricultural development \cite{Adams2021}.
The first law related to rural credit in Brazil dates back to the 1960s. Currently, the main document guiding the whole process is the Manual of Rural Credit\footnote{https://www3.bcb.gov.br/mcr}. It is an instrument with approximately 500 pages defining the following:
\begin{itemize}
    \item Stakeholders in the credit ecosystem (e.g., public banks, private banks, cooperatives, farmers\footnote{Farmers and producers are used interchangeably throughout the text});
    \item Lines of credit to cover costing (e.g., seeds and pesticides), industrialization (e.g., tractors and machinery), and commercialization (e.g., packaging and transport);
    \item Credit types;
    \item Sources of money;    
    \item Laws guiding processes;
    \item Cash flow;
    \item Responsibilities and liabilities for farmers and banks.
\end{itemize}

This provides a glimpse of the ecosystem's interconnectedness and complexity. 
For instance, banks participating in the rural credit ecosystem must employ 30\% of cash deposits (updated monthly) as rural credit. 
In cases where this amount is not met, the bank is subject to fines of up to 40\% of the non-lent money. 
In addition, banks offering rural credit are co-responsible for the proper use of the money. This involves the responsibility of performing inspections throughout the harvest and being subject to fines in case of misuse of the funds (e.g., deforestation). Regarding support offered to farmers, the organization promoting the farming activity (e.g., banks and cooperatives) can associate credit with technical support (i.e., Educative Rural Credit) as a way of mitigating risks related to yield and environmental impacts. In sum, banks need to reach and engage farmers, provide support from planning to commercialization, and inspect the proper use of land and credit, all the while respecting the environment. The amount of money lent to small farmers is up to R\$ 360.000,00 (approximately U\$ 65,500.00 in the current exchange rate), medium farmers up to R\$ 1.760.000,00 (approximately U\$ 320,000.00 in the current exchange rate), and large producers more than R\$ 1.760.000,00. These characteristics highlight how challenging is to design support systems for stakeholders in the rural credit ecosystem and how they differ from microcredit initiatives present in the literature (please refer to the Related Work section) when equitable access to wealth is a key objective.

\subsection{Context of the Study}

This field study was performed in Brazil, a country with a population of approximately 210 million people\footnote{https://www.ibge.gov.br/}, an area of 8,5 million $km^2$ (3,3 million $mi^2$), yearly agriculture production of \$140 billion dollars (approximately 21\% of gross domestic product).
Agriculture in Brazil is a highly concentrated industry, with 1\% of the farmers owning 45\% of the agricultural area \cite{Oxfam2016}. Rural credit is highly skewed to these large farms, as 75\% of rural credit (\$31B dollars) is lent to them\footnote{http://www.agricultura.gov.br/assuntos/politica-agricola/credito-rural}.
This reinforces the importance of rural credit in the country and the opportunity of using technology to ease access of the population to multiple lines of credit reaching up to large amounts of money, potentially reducing existing inequality numbers and stimulating economic development in their communities.

\begin{figure}[t]
  \centering
  \includegraphics[width=0.4\linewidth]{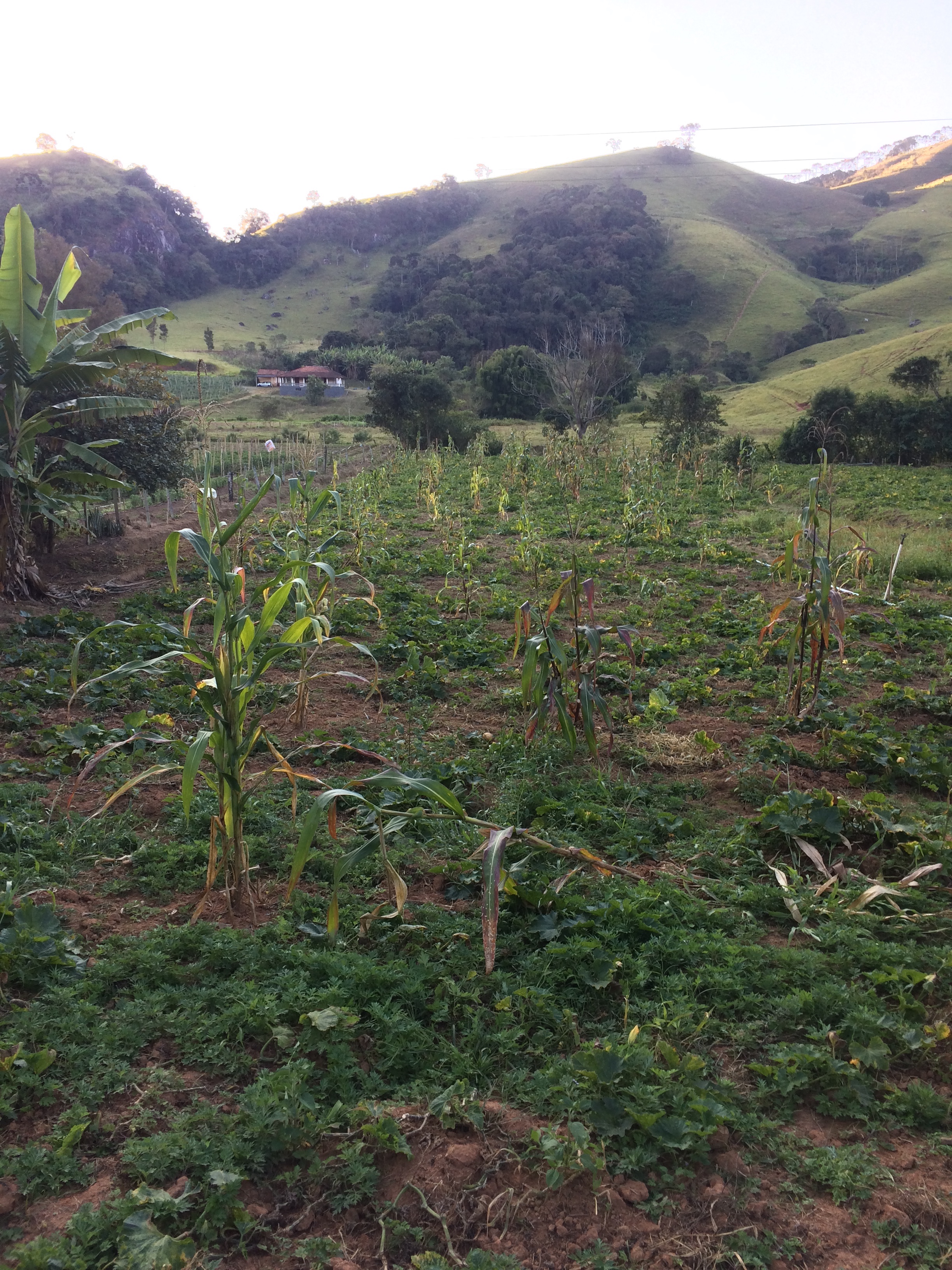}
  \caption{Photo diary picture showing one of the properties where interactions with farmers took place.}
  \Description{Photo diary picture showing one of the properties where interactions with farmers took place. It shows a corn plantation in early stages.}
  \label{fig:terreno}
\end{figure}

This work presents an in-depth six-month study involving small farmers from a rural community in Eastern State of São Paulo, Brazil (Figure \ref{fig:terreno}) 
and the experiences they have had with rural credit. 
Ethnographic methods, including participant observation, were employed aiming at understanding the challenges, opportunities, and experiences people had during the rural credit process. We used field observations, interviews, and personal experiences to understand the localized agriculture knowledge, workflows, and ways to navigate through the rural credit process, its requirements, and bureaucracies.

\subsection{Related Work}

Participant Observation has been used in multiple ethnographic studies allowing researchers to learn through different types of observation, data collection, analysis, and interpretation \cite{Spradley1980}. In the field of HCI, multiple Participant Observation studies were performed considering different populations and contexts. Examples include children with severe motor impairments \cite{Hornof2009}, business intelligence analysts \cite{Chin2009}, and therapists for children with autism \cite{Kientz2008}.

In the context of agriculture, ethnographic methods and participant observation have been applied in studies aiming for understanding the 
urban agriculture community in Australia \cite{Lyle2014}, information-sharing practices among rural users in China \cite{Oreglia2011}, and understanding the role that collaborative technologies play in skill sharing in gardening activities \cite{Maddali2020}.
 
The goal of this work builds on prior studies of credit systems (e.g., microfinance, microcredit) and farm management technologies. For example, the HCI literature considers gaps in handheld technologies for rural microfinance \cite{Parikh2006}, digitization of rural microfinance processes in India \cite{Ratan2010}, understanding of informal microcredit practices of small business owners in Brazil \cite{Candello12018}, farm management platform dealing with low connectivity and adoption challenges \cite{Vasisht2017}, and experiences and perceived obligations related to the use of instant loan platforms in India \cite{Ramesh2022}.

Having in mind aspects that result in unequal access to wealth, Marlow \cite{Marlow2005} discusses how women entrepreneurs are disadvantaged by their gender and Adams et al. \citep{Adams2021} present impacts on poor farmers as they have unequal access to loans, how initial wealth results in differential access to loans, and how they are impacted by different types of subsidies. 
Xiao-Hong \cite{Xiao-Hong2012} presents how farmers created credit cooperatives to deal with access to loan and reduced interests.
Yang et al. \cite{Yang2021} present that, in Togo, less experience, lack of membership in farmers' organization, low income on the main activity, lack of participation in training, or absence in meetings to setup of projects are the main attributes considered by financial institutions when assessing loan requests. These characteristics end up preventing them to access high value loans.
In addition, Neves et al. \citep{Neves2020} argue that rural credit in Brazil alone cannot raise the social welfare of low-income farmers and that this support should be systemic, for instance, including education and other types of support. In this work, we aim at detailing some of the complexity of the rural credit ecosystem in Brazil and identifying opportunities for supporting small farmers in a systemic way.

\subsection{Contribution Statement}

Whilst previous works mapped multiple practices from farmers around the world, there is still a valuable path to reveal experiences farms had with respect to rural credit in an ecosystem that counts on the following key differentiators: 
\begin{itemize}
    \item The government is a major player (legislation and credit offering via public banks);
    \item Private and public banks collaborating when avoiding fines and competing when offering rural credit as a product;
    \item Producers from different scales, from small properties to huge farms;
    \item From low-tech support to highly educated workers, costly machinery, and state-of-the-art technologies.
\end{itemize}

\noindent This said, the present work provides two main contributions:
\begin{enumerate}
    \item A mapping of the rural credit workflow in Brazil, its stakeholders (e.g., banks, farmers, credit specialists, credit inspectors, technicians), and their social interactions. 
    \item Implications for design and opportunities for recommender systems aiming at improving equality in loan offerings in a systemic way. The presented outcomes combine results from the Participant Observation study (bottom-up perspective) with requirements identified from Design Thinking workshops our lab ran with staff from 3 banks (top-bottom perspective). 
\end{enumerate}

This paper is structured as follows: section 2 details the method followed in the study, section 3 discusses the findings in terms of challenges and opportunities for HCI researchers, and section 4 presents the conclusions and future research directions.

\section{Method}

This study combined two methods for understanding challenges and opportunities in rural credit ecosystem in Brazil: (1) a study that took place in a community of farmers in Brazil and it was based on participatory observations of their work processes and interactions of 20 informants (bank employees and farmers); (2) design thinking workshops with teams from 3 banks, counting on 15-20 participants each.
Next, we present how Participant Observation and Design Thinking were employed to understand localized challenges connecting rural credit and farmers. 
Participant Observation is a qualitative method, with roots in ethnographic research, whose objective is to help researchers to learn the in-depth perspectives held by studied populations, interested both in knowing what those diverse perspectives are and in understanding the interplay among them \cite{Mack2005}. Design Thinking has multiple definitions and it is hard to find a single definition covering its plurality of methods, activities, and artifacts. However, some aspects related to Design Thinking converge as its aim on innovation, human needs, business success, and problem-solving exploratory practices for products/services, leveraging designers' toolkits deeply centered on human processes~\cite{Brown2008, Brown2010}\footnote{https://designthinking.ideo.com/}.
Activities covered in the Design Thinking workshops run in our lab with teams from 3 banks include As-is Scenario Map, To-be Scenario Map, Empathy Map, Hopes and Fears, Stakeholders Map, and Prioritization Grid \footnote{https://www.ibm.com/design/thinking/page/toolkit}.

Next, we provide details from the Participant Observation study. Then, in the results section, we triangulate requirements and needs identified in both Participant Observation (bottom-up) and Design Thinking workshops (top-bottom).

\subsection{Participant Observation}

\textbf{Participants:} The people contacted during the study are mainly from two different profiles: small farmers (16 people) and bank employees (4 people). Small farmers are the people from the small community that one of the authors of this paper visits regularly and owns a small property. 
A Participant Observation can be covert or overt. In a covert study, people from the studied community/group are not aware of the researcher's goal nor activity. In an overt study, people are aware that the researcher's activity/background and that they are performing a participant observation study. Different studies involve different characteristics regarding covert vs. overt. 
In this study, farmers know the researcher and know that he works with information technology, but they did not know about the study itself. This previous knowledge about the researcher prevents any valuation (positive or negative) on the researcher trying to be native.
The author in contact with farmers has been involved with agriculture for the past 5 years. This engagement also involved past interactions involving different topics. Thus, during this study, the researcher added the interest in rural credit to the list of topics to talk about.
In this sense, the study was partially covert considering the interactions and questions related to the process and experiences regarding rural credit. Other interactions involving agriculture, environment, and real estate were already in place previously to this study, in an overt way, supporting rapport building. These aspects and the already established link supported the active participation applied and the observation criteria presented in the procedure subsection. 

\textbf{Demographics:} Due to the covert aspect of the study, detailed demographics were not accessible and, by asking such type of question, it could impact the study. However, from the interactions with the participants, it is possible to highlight the following participants’ characteristics:
\begin{itemize}
\item 16 farmers and producers (from 3 different cities);
\item Farmers: 11 (as primary or secondary occupation); artisans: 3; microbrewer owners: 2;
\item Ages ranging from 38 to 68.
\end{itemize}
All farmers and 1 artisan mentioned they had in-person credit assistance. Two of the farmers informed that they tried using automated credit assistance, but they were not able to obtain credit assistance and had to go to the bank in person.

\textbf{Procedure:} The social situations that this study took place followed the five criteria for participant observation (i.e., simplicity, accessibility, unobtrusiveness, permissibleness, and frequent recurring activities), as presented in \citep{Spradley1980}. Moreover, active participation was applied considering the objective of learning about rural credit based on what members from the community have already done.

The interactions with farmers were in the form of unstructured interviews realized by convenience, in the most natural possible way. The interactions occurred during gatherings and visits to each other's properties, in public or semi-public settings. The topics in these conversations included: 
\begin{itemize}
    \item Prior good/bad experiences with rural credit;
    \item Type of credit considered;
    \item Line of credit;
    \item Blockers faced;
    \item Suggestions about doing business in the region;
    \item The whole credit process.
\end{itemize}

The visits to the community were performed similarly to how other people owning properties perform, i.e., weekly or bi-weekly. In addition, interactions were planned to occur in the most credible way, showing interest in starting in the agribusiness as a secondary occupation, which is indeed a plan for the researcher that was performing the participant observation.

In order to learn about how bankers and small farmers interact during the rural credit process, we decided to propose a feasible project and seek a small farm loan. The goal of the study was to deep dive into the process of requesting rural credit, going from obtaining the required documentation to just before signing a contract.

Regarding bank employees, they were four bank operators of the same bank brand, but from three different cities.
The bank brand is one of the key players in rural credit ecosystem in Brazil.
In preparation for these interactions, the author performing the participant observation attended different courses about the chosen production, prior to preparing a concrete project.

As a first step in selecting a small farm project, a survey on common crops and products from the region was performed. The goal was to select a product under the following restrictions:

\begin{itemize}
    \item To require management that matches the periods of visit to the small property;
    \item To fit the area of the small property (approximately 3500$m^2$ or 0.86 acre);
    \item To be profitable/economically viable.
\end{itemize}

\begin{figure}[t]
  \centering
  \includegraphics[width=0.45\linewidth]{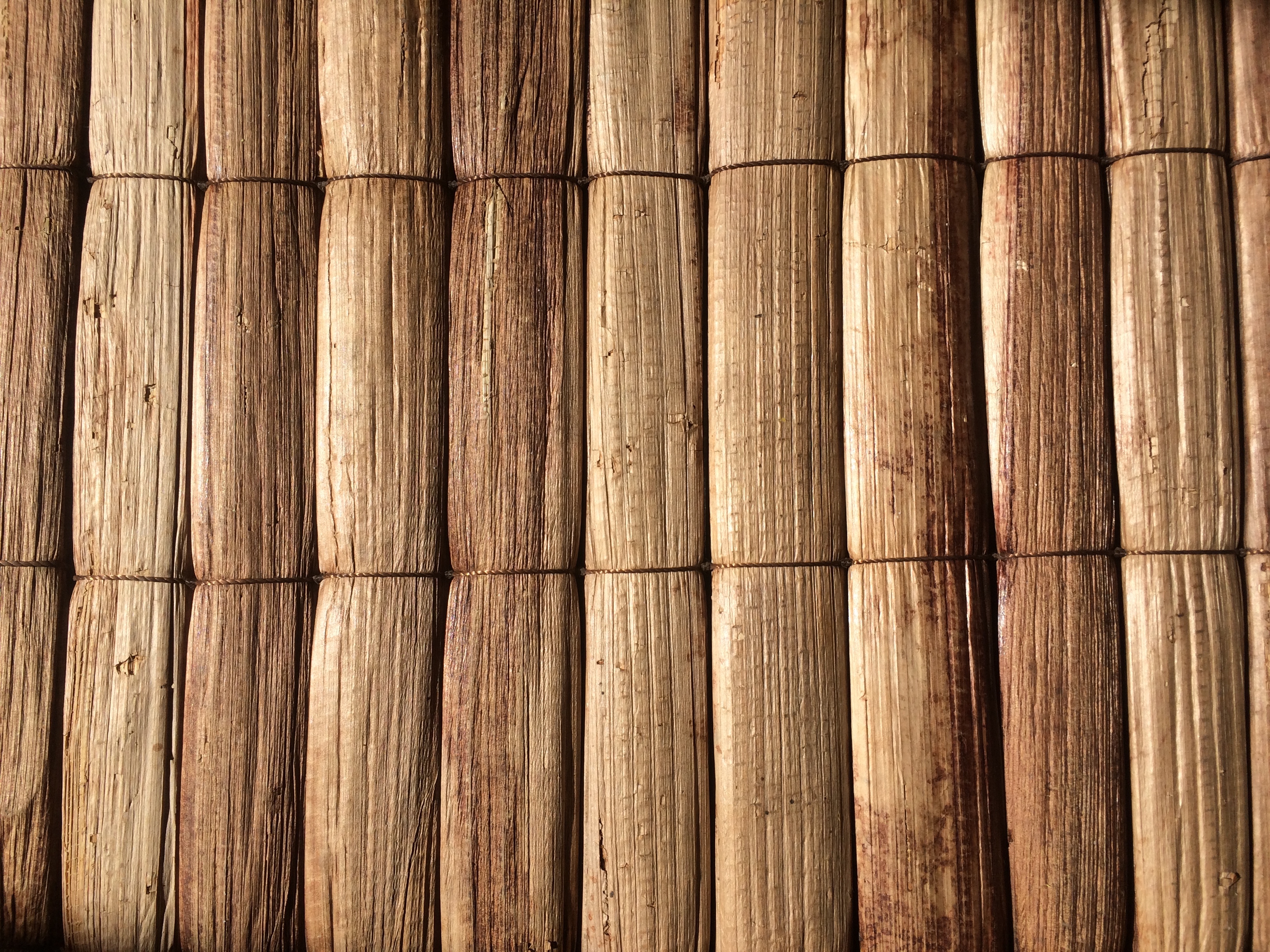}
  \caption{Photo diary picture from handcrafted mat made from banana tree straws.}
  \Description{Photo diary picture from handcrafted mat made from banana tree straws.}
  \label{fig:esteira}
\end{figure}

\begin{figure}[ht]
  \centering
  \includegraphics[width=0.45\linewidth]{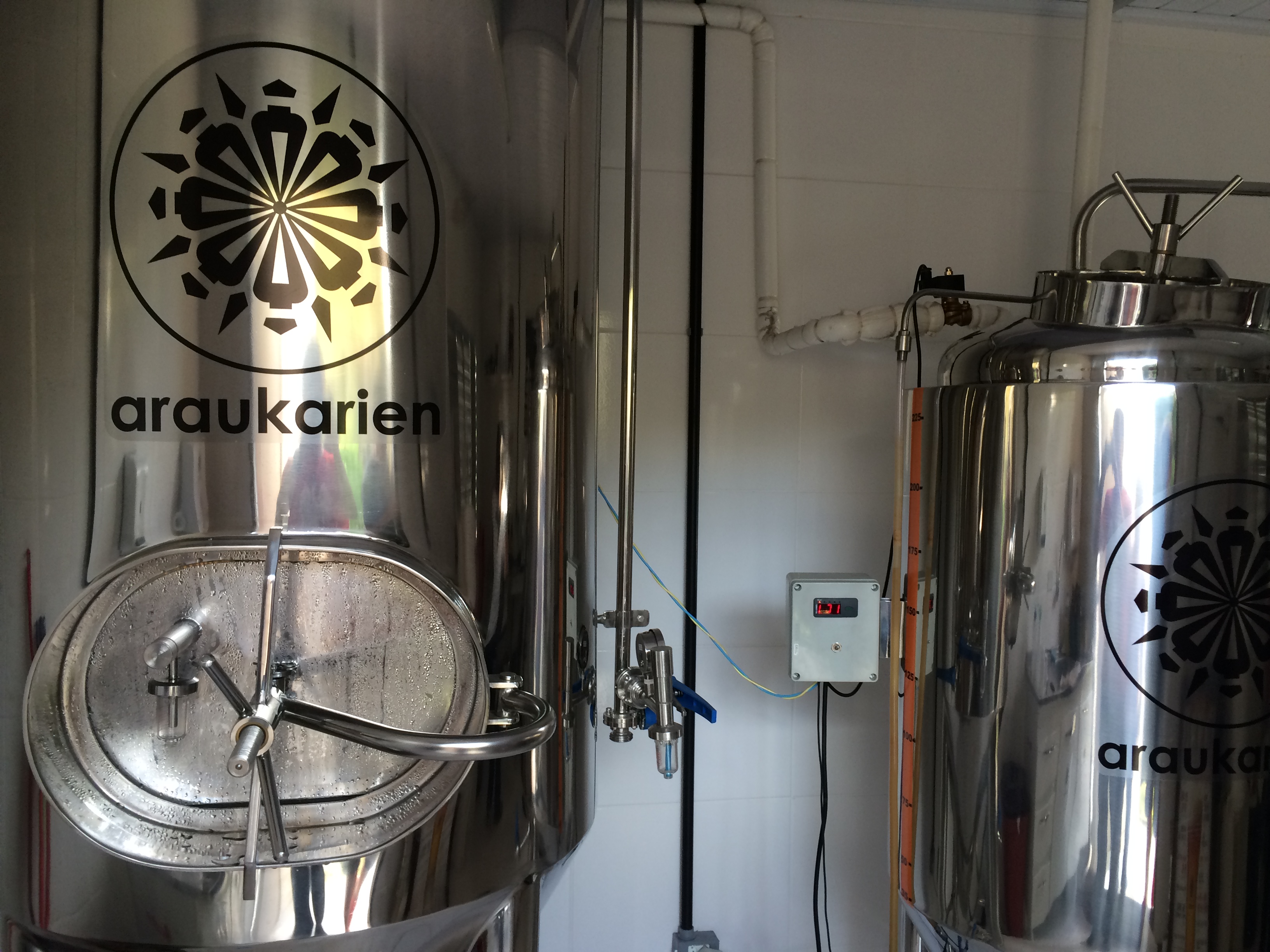}
  \caption{Photo diary picture taken during the visit to micro-brewery from the region.}
  \Description{Photo diary picture taken during the visit to microbrewery from the region.}
  \label{fig:araukarien}
\end{figure}

\begin{figure}[ht]
  \centering
  \includegraphics[width=0.3\linewidth]{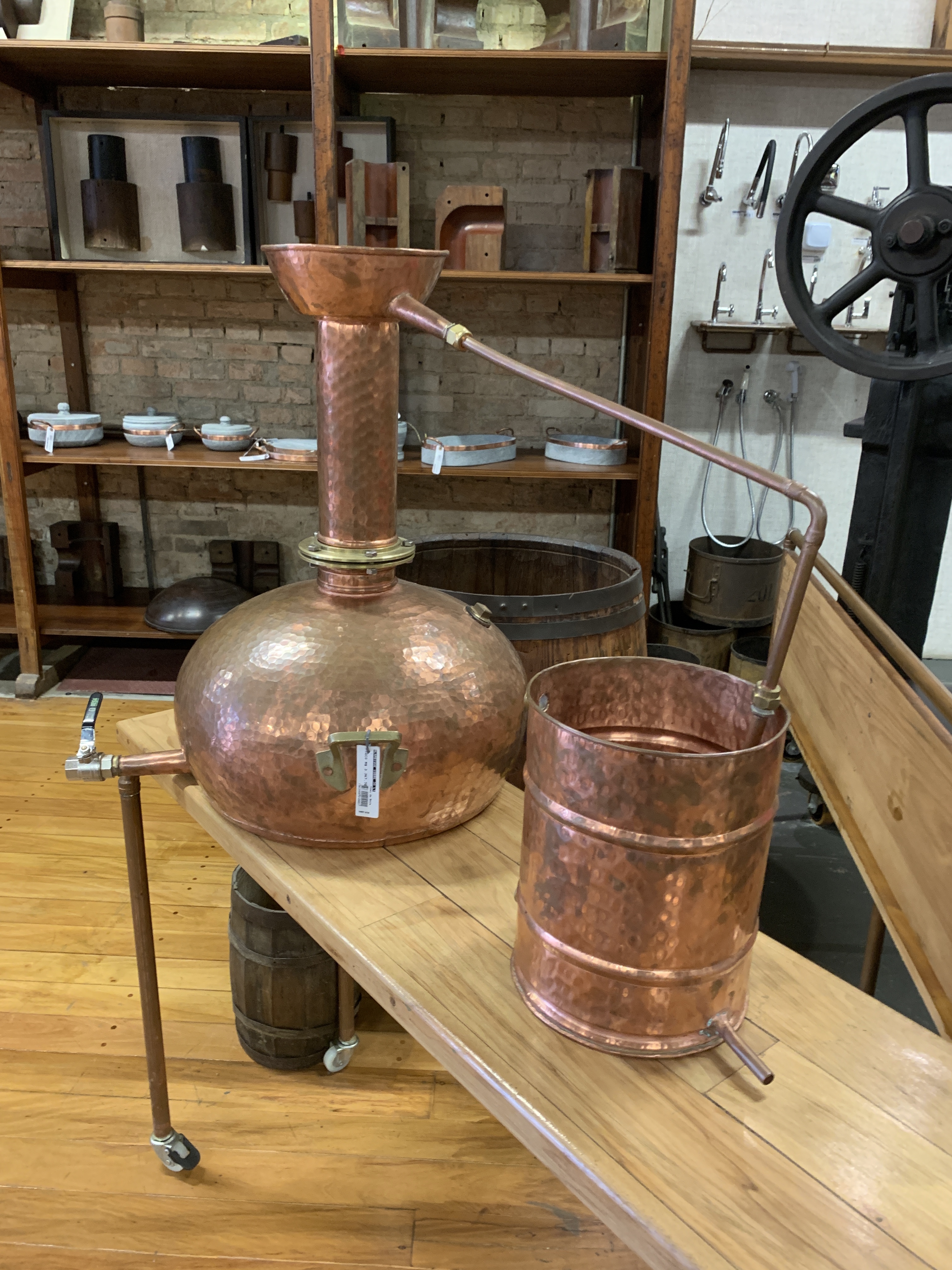}
  \caption{Photo diary picture taken in a store selling a small handcrafted copper distiller.}
  \Description{Photo diary picture taken in a store selling a small handcrafted copper distiller.}
  \label{fig:loja}
\end{figure}

The region is known for handcraft works, (Figure~\ref{fig:esteira})
crops that need cold weather (e.g., atemoya, strawberry, grape), and also agritourism (e.g., fishing weirs aside from family-owned restaurants, grape plantations from wine producers, micro-brewery). Given that the area considered is small, the initial goal was to identify a perennial/semi-perennial crop to consider. However, the identified crops would require constant management. An additional possibility was related to handcraft or fermented/distilled beverages. Thus, having in mind that the region already has breweries (Figure \ref{fig:araukarien}) 
and wineries, the opportunity was to propose something in the area of spirits. Once the production was selected, the researcher performing participant observation started to enroll in courses on the selected matter, research about cost of production, equipment (Figure \ref{fig:loja}), etc. All of this supported the elaboration of a project meant to be economically viable, considering all the informed restrictions.

As part of the project, there were multiple conversations with bank employees about rural credit loan. Conversations took place with four bank operators of the same bank brand, but from three different cities. The main goal here was to interact with bank employees in the most natural and credible way as well. The interactions were towards obtaining more information and pointers about rural credit and the required documents. These employees were all in the client-facing credit operations.

\begin{figure}[t]
  \centering
  \includegraphics[width=0.45\linewidth]{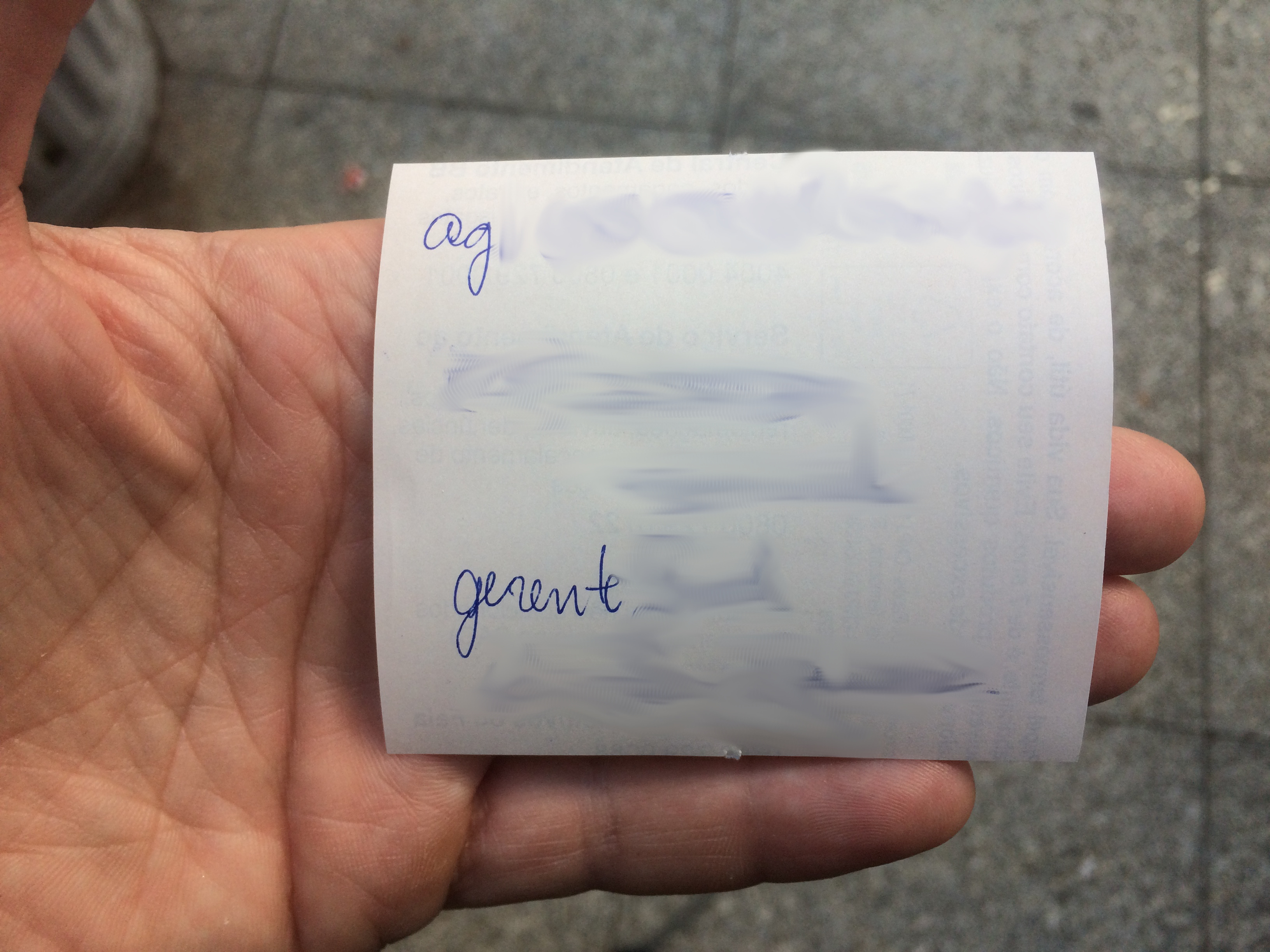}
  \caption{Photo diary picture from a bank employee's handwritten note suggesting a contact in a different agency. The employee mentioned that in that agency it was hard to find anyone knowing about rural credit and that I should look for agencies in another city; blur applied to note due to privacy issues. The words in Portuguese reffer to agency and manager.}
  \Description{Photo diary picture from a bank employee's handwritten note suggesting a contact in a different agency. The employee mentioned that in that agency it was hard to find anyone knowing about rural credit and that I should look for agencies in another city; blur applied to note due to privacy issues. The words in Portuguese reffer to agency and manager.}
  \label{fig:telefone}
\end{figure}

\textbf{Materials:} In terms of materials used and how the study was documented, notes were taken post-interactions (in a covert way) and, when possible, pictures were taken to register places and outcomes of interaction with stakeholders in a form of a photo diary (e.g., Figure \ref{fig:telefone}).
No pictures were taken from people to respect privacy and the validity of the study. The goal of using the photo diary was to support recalling the conversations’ content in post-interaction notes and during result analysis. 

\textbf{Analysis:} The result analysis considered processing the photo diary, consolidated field notes, and stakeholders' quotes and moods. 
Then, experiences with rural credit and opportunities for HCI practitioners were mapped in an end-to-end workflow of rural credit in Brazil. The workflow details steps, stakeholders, social interactions, design implications, and opportunities for researchers designing/developing recommender systems in the Brazilian rural credit ecosystem aiming at providing equal access. Finally, we also summarize the characteristics of the Brazilian rural credit ecosystem in a mind map (Figure \ref{fig:mind_map}) by connecting the main entities, terms, processes, stakeholders, and highlighting (in bold edges) opportunities for designing recommender systems providing equal access to small farmers.

\subsection{Design Thinking Workshops}


The Design Thinking workshops were organized to bring specialists from banks to detail the challenges they face in the rural credit process and to brainstorm solutions. From our side, we also had scientists and technical specialists in these sessions to understand the problems and opportunities in this realm. Each workshop was a one-day event comprising multiple activities such as Stakeholder Mapping, Scenario Mapping (As-is and To-be), Empathy map, and Prioritization Grid. Each workshop involved 15-20 people, including our staff and bank participants.

\section{Results} 

\begin{figure*}[ht!]
  \centering
  \includegraphics[width=\textwidth]{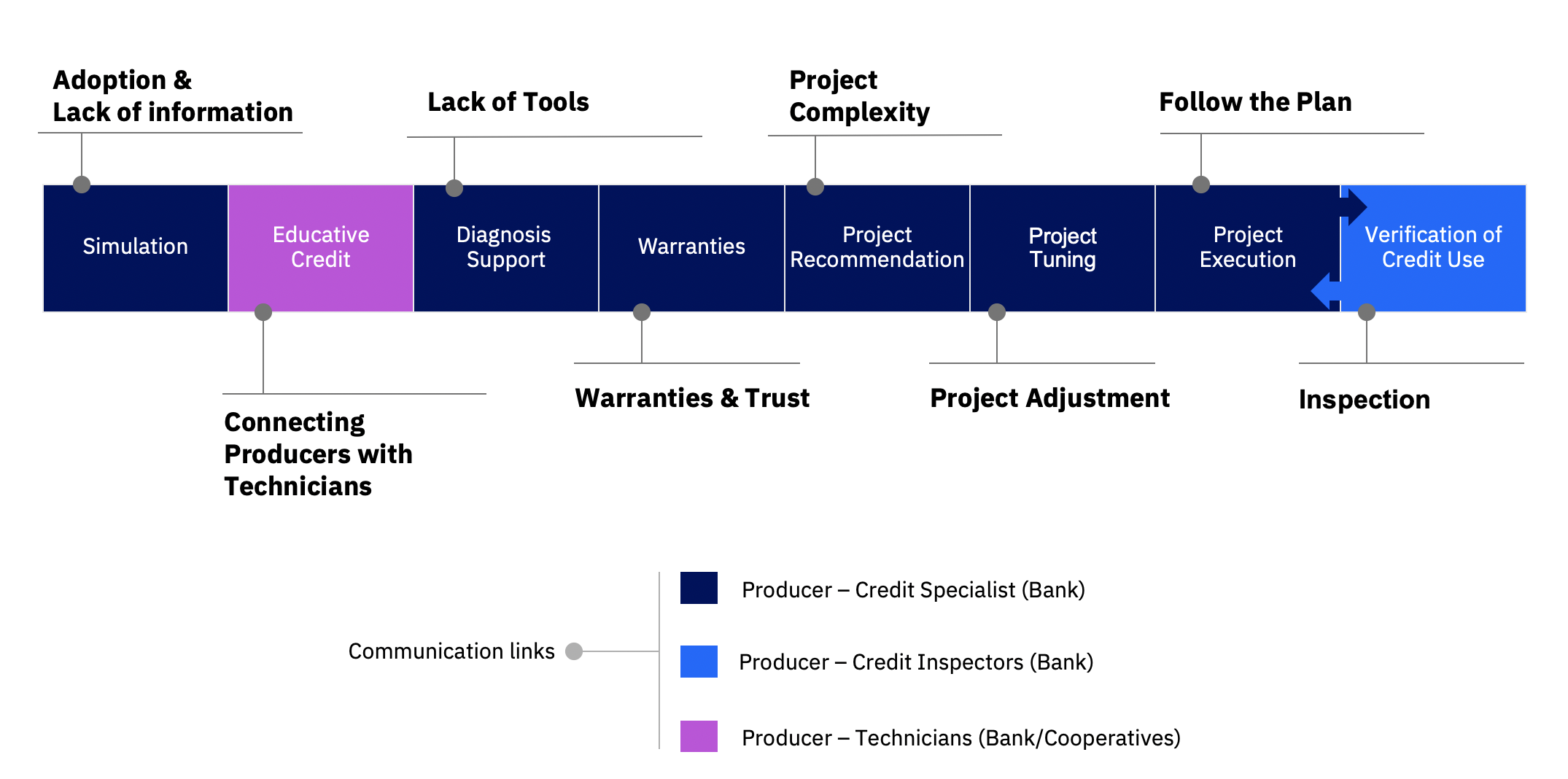}
  \caption{Rural credit workflow in Brazil, stakeholders, social interactions, and design implications.}
  \Description{Rural credit workflow in Brazil, stakeholders, social interactions, and design implications.}
  \label{fig:workflow}
\end{figure*}

In this section, we present the identified nuances of rural credit workflow in Brazil (Figure \ref{fig:workflow}) and the implications for HCI research, from a design perspective, and opportunities for fair recommender systems technologies that could change the \textit{status quo}. The identified workflow encompasses: 
\begin{itemize}
    \item Simulation of credit;
    \item Optional technical (e.g., agronomist, veterinarian) support as part of educative credit;
    \item Diagnosis support for the farmer to improve production;
    \item Warranties and associated documentation required;
    \item Project setup and fine-tuning;
    \item A loop involving project execution and verification of credit use until the end of the project.
\end{itemize}

Figure \ref{fig:workflow} also highlights that most of the social interactions occur between the producer and the credit specialist, followed by the producer-credit inspector and the producer-technician interactions. 
In addition to the results presented from the participant observation study, we also triangulate these outcomes with insights from Design Thinking workshops our lab run with bank staff.
In the period this work took place, our lab interacted with 3 banks interested in brainstorming solutions for credit and related products/services. The Design Thinking sessions were run for a single bank at a time; bank names are omitted due to confidentiality terms. The rationale for combining those two sources of information is to triangulate bottom-up requirements provided by the farmers with the top-bottom requirements provided by the bank staff.
The next subsections present each of the steps from rural credit workflow and respective challenges and opportunities for equal access to wealth via high-value loans with low interest rates.

\subsection{Rural Credit Simulation}

Often, the first step for farmers requesting rural credit is the simulation, including payment terms, lines of credit, documentation, among others. It can be done on bank websites or in person by talking to bank employees. Thus, this first step identified is connected to the design of tools and the adoption of such technologies that clearly communicate rural credit terms and caveats. 

In rural Brazil, 61\% of producers use smartphones and WhatsApp® is the main communication channel, used by 96\% of farmers with internet access\footnote{https://www.embrapa.br/visao/}. One of the probable root causes is that, in Brazil, multiple telecommunication companies provide plans in which such a service does not impact the monthly quota of data consumption. Hence, this is one platform with potential for easing the adoption of credit simulation technologies. 

Although, when asking a participant about how to perform credit simulation and any existing mobile app, a farmer reported: ``I tried using an app for recommendations about crop management, but it kept asking me the same questions over and over'', showing that a bug tarnished his experience with this chatbot. Having the adoption of technologies in mind, HCI researchers could leverage existing communication channels to speed up the adoption of decision-making technologies. For instance, connecting chatbots with WhatsApp® users/public groups, bridging the gap between established communication channels and support technologies. 
 
In this step, it is also key to identify whether farmers comply with all the requirements. However, it was identified that this process is cumbersome and information is often spread in the sense that few bank employees know about rural credit; please see Figure \ref{fig:telefone} for a case in which a bank employee indicated a different agency that maybe they would know the required documents. Two farmers mentioned that in the past seasons it was easier to obtain rural credit. They reported that the process’ bureaucracy increased and it is becoming more and more difficult to obtain credit. One participant said: ``In the last years, I've just had to show some of my handwritten notes about the production that the bank accepted. Now, I have to prepare documentation about the property, production, and warranties that were not needed in the past.'' 

Beyond the bureaucracy, during social interactions with bank employees, it was hard to get information in person. Only after visiting 4 different agencies, it was possible to identify the documents the bank required. In the first visited agency, the bank employee was not able to inform the list of documents required and tried to connect us to another credit specialist (Figure \ref{fig:telefone}). After this contact, this happened one more time, showing that, even for the client-facing bank employees, the lack of information about the process happens. After talking to credit specialists, it was possible to obtain the list of required documents. In terms of technology for automating this process, this step could be structured as a decision tree covering, for instance, production in the past year, warranties, amount of money, payment terms, etc. 

Regarding design implications, a step-by-step form (auto-complete enabled) or a chatbot could convert the steps on this decision tree into a user interface that could result in a good user experience from the first step.

In a Design Thinking session with a bank, it was clear that it is necessary to provide automatic credit for producers in some cases. If the bank has documents and information about the farmer, income, and rural production, it is possible to approve credit automatically. However, in some cases, the farmer needs to go to the bank for a new credit. Another possibility involves providing self-education content related to this pre-approval credit to the farmer.

\subsection{Educative Credit}
 
Rural credit in Brazil has different types and different services offered as part of the loan. One considers technical support for the farmer (i.e., Educative Rural Credit). The goal of this credit type is to mitigate risks for the producer and for the bank, since a technician support (e.g., by an agronomist or veterinarian, depending on the production) is offered as part of the planning steps. Under the lens of decision-making support systems, this type of credit creates an opportunity for providing educational/technical content to small farmers, from planning to harvesting. For instance, supporting the best time window to plant/harvest, in a data-driven way and what types of pesticides to use, in an environment-respecting way.

The design implications here are mostly connected to learning systems and lifelong learning, exploring multiple modalities and connectivity restrictions that may apply in a country with challenging contexts of diversity, (digital) literacy, and connectivity. In such a diverse socioeconomic context, support/educational content might be transformative for some people when quality content meets the need to know.

In Design Thinking sessions with banks, it was reported that small farmers usually do not have enough knowledge about historical information involving soil characteristics, climate, yield on other farms in the region, among other rich datasets. Often, they repeat the same maintenance and crop without knowing that they could be more successful with a different crop, especially when mitigating risks related to price oscillations. 

\subsection{Diagnosis Support}

Small farmers usually plan, manage, and harvest based on tacit knowledge. Diagnosis of field is hardly considered in rural credit projects due to its associated costs (e.g., soil tests). In Brazil, there are producers that still use fire as a method for preparing the soil, which has potential environmental impacts. In this sense, low-cost technologies supporting the diagnosis and soil preparation are key for supporting the understanding of the underlying factors that might increase productivity/quality, respecting the environment. 

The implications for HCI researchers in this step are related to the use of sensors and low-cost devices. Diagnosis support might include characteristics from soil, seeds, precipitation, satellite images, weather forecast, among others. Thus, beyond the use of sensors, accessible information visualization could play a key role in supporting the understanding of characteristics of the field.

In Design Thinking sessions with banks, a need emerged for a solution employing easy to use geo-referenced management system, providing farmer recommendations and bank information that the farmer is employing credit responsibly. Such a solution has the potential to facilitate further processes due to the seamless compliance information provided throughout the harvest cycle. Beyond diagnosis support for the crop itself, bank staff also highlighted that producers often need support to manage future income, combine profitable crops, identify best-selling moment, and perform scenario analysis/simulation.

\subsection{Warranties}

As part of the credit, small farmers must present a warranty to the bank prior to obtaining the loan. For small farmers with good credit scores, the risk evaluation is straightforwardly performed by banks. However, for newcomers or digitally excluded producers, the lack of history poses a challenge for both ends, farmers and banks. In this sense, technologies for finding analogous profiles (e.g., by crop, region, field area) and designs exposing how such credit was performed (explainability\footnote{Explainability or Explainable Artificial Intelligence (XAI) 
is a research field that aims to turn AI results and models more understandable to humans\cite{adadi2018peeking}.}) could support this step. These analogous profiles could also be a data source for recommending warranties usually considered, including project templates, project recommendations, and insurance pricing.

Banks employees and producers are the main stakeholders in rural credit workflow presented, hence, trust in this relationship is fundamental. However, in two occasions, informants mentioned that their banks often offer products with the rural credit as part of the warranties, e.g., life insurance, disability insurance, among others. One participant said: ``If you contract [a life] insurance, you'll get the credit right away.'' Other participant mentioned: ``At some point, I had 3 different life insurances to pay and I’d lost track of how much I was paying for it''; the later fact is connected to the automatic renewal for producers contracting rural credit yearly. 

The design implications here consider clearly communicating the renewal terms and conditions in an accessible way.

\subsection{Project Recommendation}

The complexity involving rural credit was highlighted in the simulation step when farmers are trying out different amounts of money, payment terms, and lines of credit. However, creating the project also poses and additional challenge due to the technical requirements, including the details about money usage in each of the planned activities. In this sense, the number of details required for newcomers might create a barrier that the researcher performing the participant observation faced when creating the first project. Thus, one opportunity for recommender systems includes providing project templates, based on analogous successful past projects. For instance, involving similar crops, similar regions, similar weather, fields with similar characteristics. These analogous projects could also be a valuable education material for farmers getting in touch with the rural credit for the first time. 

The main design implication in such a feature includes communicating with users in a way to obtain enough information for finding an analogous project in a privacy-respecting and accessible way.   

In Design Thinking workshops with banks, it was discussed the possibility of using climate prediction, sensors, and drones/satellite images to assess and recommend equipment to purchase for lines of credit related to industrialization and commercialization. This way, project recommendation can go beyond the current farm capabilities and increase productivity.

\subsection{Project Tuning}

This step includes adjustments that might be needed after the producer submits the documentation and the project to the bank. In this step, the credit specialists will analyze the proposed planning and requests adjustments in case of any discrepancy according to bank metrics. In such iterations, minor changes that can delay the loan approval in days/weeks, could impact planting date and, consequently, yield. Bearing in mind opportunities for decision-making support technologies, it is straightforward to think about preventing these discrepancies to occur in the project creation/recommendation by providing user feedback and education material for any outlier identified, in a way that credit specialists could support producers in a more agile way.  

Design implications involve, for instance, providing user feedback about these outliers, based on analogous projects considering weather, crop, region, etc. 

\subsection{Project Execution}

After minor adjustments and approval, the project execution loop starts. In this step, the challenge is to guarantee that the plan is being followed. Hence, differently from the previous steps that occur once per project, this step might get as complicated as daily managements activities, depending on the production considered. Thus, the balance between too detailed and too vague, considering farmers' tacit knowledge, seems to be the ultimate goal. In sum, providing personalized support.

The implications for designing include connecting to external data sources (e.g., sensors, weather services) to support farmers to perform well-informed decisions. Accessible charts/dashboards could be provided based on interaction history and farmer engagement with services daily (e.g., weather, precipitation, temperature). 
Big producers count on teams of high skilled professionals to perform continuous data analysis. However, supporting small producers may be the real challenge/opportunity, for instance, mobile technologies easing crop inspection activities and project execution report. For instance, taking pictures of the crop and performing computer vision algorithms to identify proper crop development, pest, and diseases. 
Moreover, low tech support materials as precipitation charts, weather forecast, task list, paper-based soil testes, week schedule to be printed and attached in visible areas can increase awareness of the plan and support all stakeholders on following the plan. 

\begin{figure*}[ht!]
  \centering
  \includegraphics[width=1\textwidth]{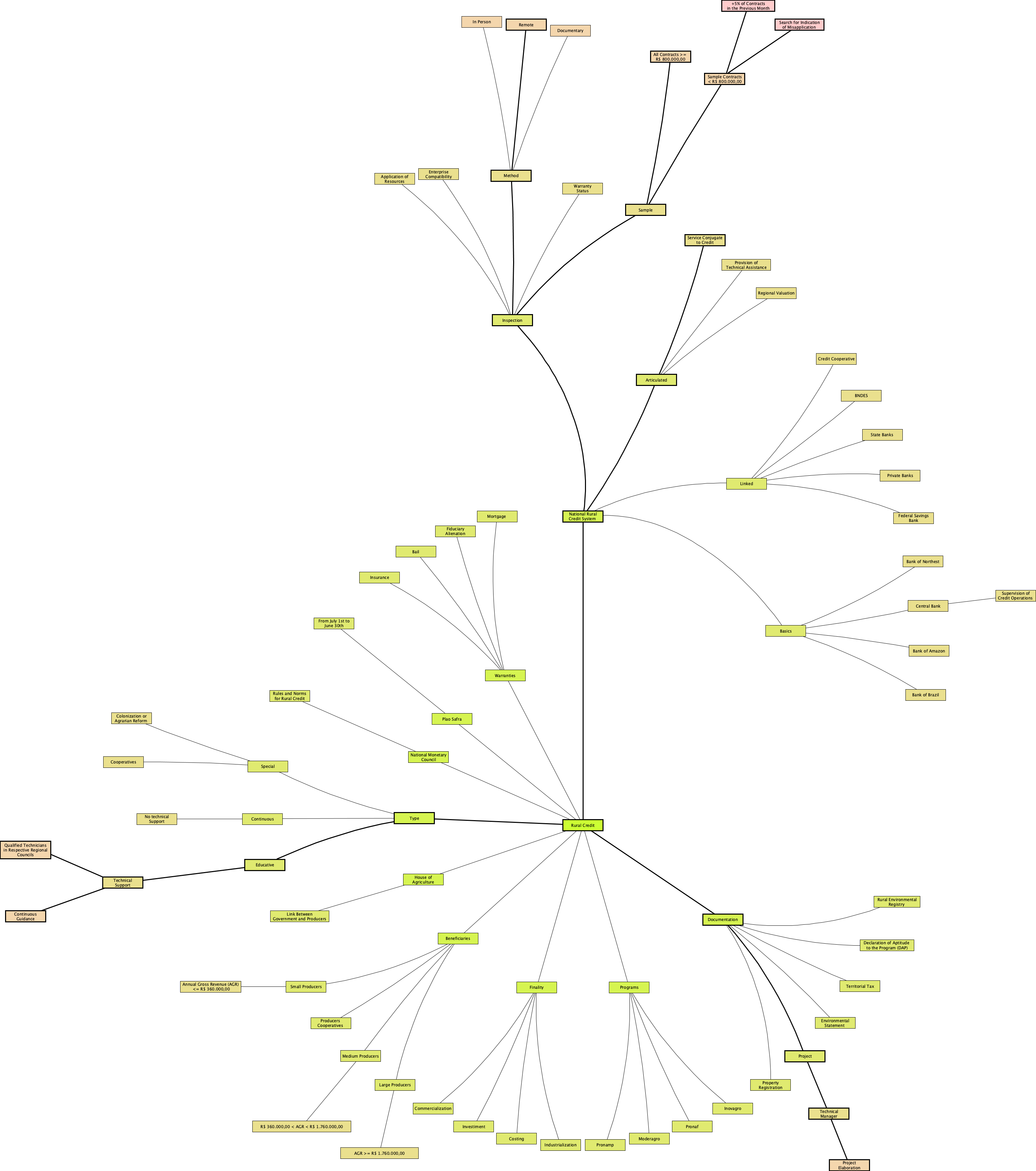}
  \caption{Mind map highlighting the main entities in the Rural Credit ecosystem in Brazil. Bold edges and nodes highlight opportunities for designing recommender systems.}
  \Description{Mind map for the Rural Credit ecosystem in Brazil. Edges and nodes in bold highlight opportunities for designing recommender systems.}
  \label{fig:mind_map}
\end{figure*}

Results from Design Thinking workshops point to the need of providing easy ways for the farmer to easily verify cash flow, credit limit, pricing, and to improve production management. 

\subsection{Verification of Credit Use}

As presented before, in the ecosystem of rural credit in Brazil, banks and credit cooperatives are also liable for the proper use of the credit and for the land where the crop will take place. These stakeholders must verify that the credit is being used for the planned crop and that it is not damaging the environment, just to name a few responsibilities. In this sense, decision-making support technologies can help small farmers to follow the plan and also remote sensing can be applied to verify the proper use of the land/credit. Moreover, Educative Rural Credit and technician support could close this loop. For instance, a producer could ask for technician support by sending a geo-referenced picture of a pest or plant with disease; this would help seamless verification of credit use associated with technician inspection.

The last step mapped in the rural credit workflow is the renewal in-between seasons. One participant mentioned that renewing was easy: ``I used the rural credit 3 years in a row. And it was easy to renew it. At the end of the season, the money was there again''. However, the same participant informed that: ``At some point it was difficult for me to fall asleep due to the multiple payments installments I had''. In this aspect, the Educative Rural Credit could support small farmers not only on how to renew credit, but also with accounting when dealing with multiple loans. A cohesive view on loans and possibilities on how to improve productivity (via commercialization and industrialization) poses new possibilities for finance and education support as well.

In a Design Thinking workshop, bank staff mentioned that banks have a huge challenge to assert that the farmer is using credit for the production proposed in the project. This activity was usually performed by credit inspectors, in person, but currently remote sensing and image recognition are used to enhance the analysis and prevent frauds in a more scalable way. Finally, at the renewal, banks value proactive/automatic actions to support credit renewals based on the clients' history of credit and production. 

\section{Conclusion}

Differently from microcredit offerings, rural credit in Brazil is a line of credit present in multiple bank brands aiming at all kinds of production, from small farmers up to huge producers.
For small farmers it has an important socioeconomic role as it provides financial support for producers in costing, industrialization, and commercialization. 
While rural credit counts on annual fees ranging from 2.5 to 4.5\%, banks in Brazil offer monthly fees of approximately 2.0\% for other lines of credit. In addition, banks are responsible for maintaining a credit flow according to cash deposits, which presents a challenge of offering more credit, reducing risk, dealing with adoption of recommendation technologies, and supporting small farmers to grow via Educative Rural Credit and, last but not least, respecting the environment. 

Responsible Innovation is often described in terms of three main dimensions: (1) avoid harm, (2) 'do good', and (3) support ethical governance to promote the former two dimensions \cite{Voegtlin2017}. Initiatives to promote access to wealth and social mobility initiatives count on practices and regulations to avoid harm, however, there is a gap between planned outcomes and actual results and access to wealth in the case of rural credit in Brazil (i.e., 'do good' in a socio-economical way). Bureaucracy, lack of support to small farmers in terms of technology, project preparation, and guidance, are some of the key barriers between small farmers and access to rural credit in Brazil. 

In this work, we presented how participant observation was employed as a way of identifying \textit{localized challenges}, inherent to agriculture, and how fair recommender systems (from the design perspective) can support multiple stages of the rural credit workflow. The initial objective of the study was to experience the rural credit process up to just before the contract signing. The blocker experienced by the researcher performing the participant observation was to present the history of production and to obtain the documentation proving previous activity in agribusiness. These aspects resonate with results from the literature about blockers small farmers may face \cite{Yang2021}. In a complementary way, different outcomes from Design Thinking workshops were presented to depict a top-down perspective (from bank to producers) on requirements, interests, and goals. 

This study presented challenges faced by farmers in a small community in Eastern State of São Paulo, Brazil, in multiple rural credit stages as request, settle payments, and renewal. Agriculture is very localized due to all its impacting variables. For HCI researchers studying agriculture domain for the first time, the presented findings can be a starting point for identifying practices and pain points that multiple stakeholders face. More importantly, showing that small farmers should also have access to different lines of credit, beyond microcredit.

The mind map presented in Figure \ref{fig:mind_map} summarizes multiple aspects described and discussed throughout this paper. Its goal is to show the connections among terms, stakeholders, and processes in a way that HCI practitioners can grasp the overall ecosystem complexity, impacts on unequal access to credit, and identify opportunities as well. Bold edges highlight these opportunities encompassing design and development of recommender systems, as detailed in previous sections.

\begin{figure}[t]
  \centering
  \includegraphics[width=0.45\linewidth]{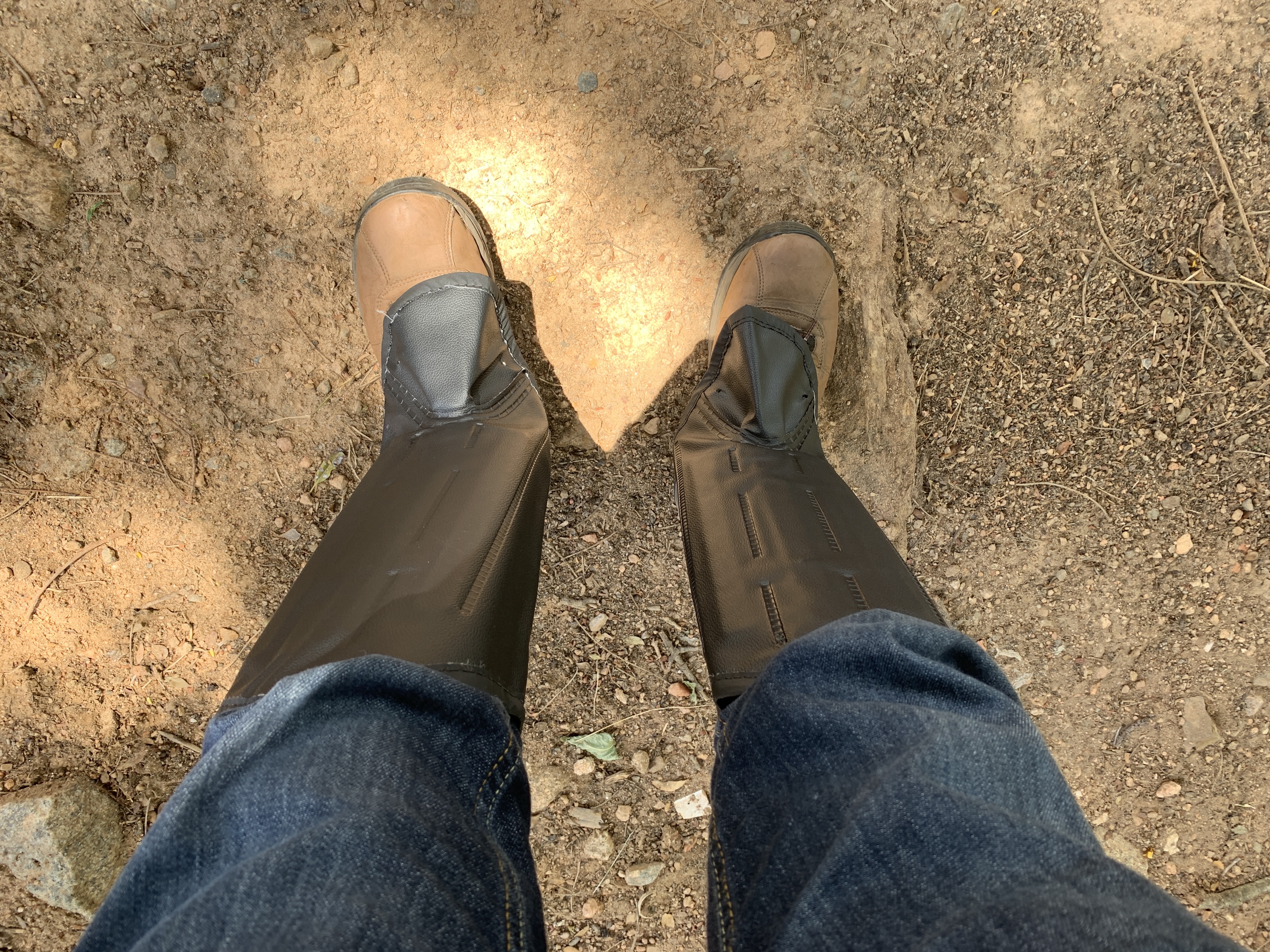}
  \caption{Photo diary photo showing protective shin guards used to protect from snake bites used during a visit to a coffee farm.}
  \Description{Protective shin guards used to protect from snake bites used during a visit to a coffee farm.}
  \label{fig:caneleira}
\end{figure}

After connecting with multiple banks and agribusiness companies, people in this domain usually mention two distinct types of interactions with agriculture: \textit{inside the fences vs. outside the fences}. The idea behind this differentiation is to emphasize the importance of getting closer to the subject matter and the real challenges faced by the people. Figure \ref{fig:caneleira} shows the safety equipment used against snake bites during one of those interactions with clients. Knowing that HCI literature advocates using Ethnographic methods, User-Centered and Participatory Design in such scenario, we emphasize the value of these approaches based on this study, especially in agriculture.

In this sense, this study pointed out promising directions for actionable results in the rural credit ecosystem taking into account multiple stakeholders, e.g., public banks, private banks, cooperatives, input companies, and developers of recommender systems. 

This paper contributed with reflection on challenges and opportunities in providing small farmers access to a line of credit with potential to impact wealth (i.e., rural credit).
In addition, we provided insights on the design of recommender systems aiming at equal access to loans based on the experiences reported by informants and by the situations faced during the study.
We also detail the rural credit ecosystem in Brazil, depicting stakeholders and their interactions, as well technologies that could be employed in a systematic way to improve decision-making support. This paper also triangulates (bottom-up) requirements emerging from interaction with farmers with (top-bottom) requirements elicited during Design Thinking sessions with banks' staffs. This combination grounded the paths we proposed regarding recommender systems aiming at increasing loan signings, promoting better bank-farmer interactions, improving farmer experience in the whole process, and promoting fair access to loans.

We believe that recommender systems in rural credit ecosystem have different purposes as easing the access of the population to multiple lines of credit reaching up to large amounts of money, potentially reducing existing inequality numbers and stimulating economic development. In addition, due to heterogeneity of profiles, non-negligible group of small farmers may not benefit from access to rural credit, which potentially impacts their economic ascension. We also identified that such recommender system can be beneficial for producers and banks. On the one hand, it might increase access to credit, potentially reducing existing inequality numbers and stimulating economic and educational development. On the other hand, it might improve bank-client relationship and overall economy. In sum, \textbf{access, equity, and justice}.

The main lesson learned worth sharing with designers and HCI researchers is that, due to its localized nature and heterogeneous stakeholders’ profiles, it would be difficult to obtain a user interface to fit all stakeholders’ needs around the globe. Hence, when dealing with agriculture, we advocate that designers and HCI researchers should focus on employing methods to tackle localized design challenges (e.g., Ethnographic methods, User-Centered Design, Participatory Design). The research related to agribusiness stakeholders require in-depth study about their activities and the ecosystem they are embedded in. Agriculture counts on specific, localized challenges requiring HCI researchers to go the field to observe, survey, interview, understand pain points, listen to needs, and envision solutions and technologies that materialize these solutions, \textit{inside the fences}.

Next steps of this research involve exploring recommender systems technologies aiming at equal access, providing explainable outcomes about Machine Learning models, supporting credit specialists, farmers, and rural credit newcomers to benefit and to be part of the rural credit ecosystem and having equal access to wealth.


\bibliographystyle{ACM-Reference-Format}
\bibliography{sample-sigconf}





\end{document}